\documentclass[fleqn,12pt,table]{wlscirep}
\usepackage[utf8]{inputenc}
\usepackage[T1]{fontenc}
\usepackage{makecell}

\definecolor{Gray}{rgb}{.9,.9,.9}

\title{Is it getting harder to make a hit? Evidence from 65 years of US music chart history}

\author[a]{Marta Ewa Lech}
\author[a,b]{Sune Lehmann}
\author[a,b]{Jonas L. Juul}

\affil[a]{Department of Applied Mathematics and Computer Science, Technical University of Denmark, DK-2800 Kgs. Lyngby, Denmark}
\affil[b]{Copenhagen Center for Social Data Science, University of Copenhagen, DK-1353 Kgs. København K, Denmark}

\begin{abstract}
Since the creation of the Billboard Hot 100 music chart in 1958, the chart has been a window into the music consumption of Americans. Which songs succeed on the chart is decided by consumption volumes, which can be affected by consumer music taste, and other factors such as advertisement budgets, airplay time, the specifics of ranking algorithms, and more. Since its introduction, the chart has documented music consumerism through eras of globalization, economic growth, and the emergence of new technologies for music listening. In recent years, musicians and other hitmakers have voiced their worry that the music world is changing: Many claim that it is getting harder to make a hit but until now, the claims have not been backed using chart data. Here we show that the dynamics of the Billboard Hot 100 chart have changed significantly since the chart's founding in 1958, and in particular in the past 15 years. Whereas most songs spend less time on the chart now than songs did in the past, we show that top-1 songs have tripled their chart lifetime since the 1960s, the highest-ranked songs maintain their positions for far longer than previously, and the lowest-ranked songs are replaced more frequently than ever. At the same time, who occupies the chart has also changed over the years: In recent years, fewer new artists make it into the chart and more positions are occupied by established hit makers. Finally, investigating how song chart trajectories have changed over time, we show that historical song trajectories cluster into clear trajectory archetypes characteristic of the time period they were part of. The results are interesting in the context of collective attention: Whereas recent studies have documented that other cultural products such as books, news, and movies fade in popularity quicker in recent years, music hits seem to last longer now than in the past. 
\end{abstract}
\begin{document}
	
	\flushbottom
	\maketitle
	\thispagestyle{empty}

	\section*{Introduction}
	The Billboard Hot 100 music chart has quantified American music consumption since the chart's creation in 1958. 
	For more than half a century, the Billboard Hot 100 has been a primary authority for gauging the significance and popularity of songs and musical artists \cite{chartlegend}. 
	Society has changed a lot since the chart's inception in the Fabulous Fifties: 
	The world has become increasingly globalized, and new technologies have emerged including CDs, DVDs, and streaming services for music distribution, MTV and YouTube for music video broadcasting, and social media and the internet that allow faster-than-ever communication between fans. 
	
	In recent years, the technological acceleration has been particularly pronounced. 
	The total number of music streams were estimated at  2.17 trillion worldwide in 2020 \cite{streamingstats}, and experts report that more than $100\,000$ new songs are uploaded to the streaming service Spotify every day \cite{spotifydaily}.  
	Ryan Tedder, a successful producer and songwriter, recently said in an interview with the BBC that the popularity and capabilities of streaming services have made it harder for artists to create new hits \cite{teddersongs}. 
	First, newly released songs are competing against every other song released in the past 70 years. Apart from that, with tens of thousands of songs uploaded to Spotify every day, the industry has become more competitive than ever. 
	This is the perspective of an artist and creator, but can we quantify how song and artist performance on music charts changed in recent decades? 
	
	In this paper, we use the Billboard Hot 100 historical records to examine how the success of artists and songs have changed in recent years. 
	To this end, we investigated the changes in the dynamics of the Billboard Hot 100 weekly music charts over time. 
	The chart combines album sales, airplay, and streaming to identify the most popular musical tracks each week.  
	We analyze different aspects of music rankings, such as lifetimes of songs on the chart, weekly position changes of chart songs, as well as the artists' song counts on the chart. 
	By comparing these features over time, we discover trends in how the song performance changed over the last 6 decades. 
	Among other things we find greater inequality in song performance on the chart in recent years: 
	Nowadays, the most popular songs stay popular for longer than ever, whereas the least successful songs on the chart are replaced quicker than in the past. 
	Today's chart also include fewer new artists and more songs by established hitmakers, and recent years has seen a surge in the number of collaborations and features on the chart. 
	Furthermore, we use chart position sequences to identify song trajectory archetypes. 
	We find that each song archetype has been prevalent on the chart in certain well-defined time periods, indicating clear changes in chart dynamics over time. 
	
	Our paper is structured as follows. 
	Following a literature review, we present our data and methodologies in the section Materials and Methods. 
	We then proceed to present our results and conclude the paper by discussing our findings and putting them into the context of the broader literature.
	
	\section*{Related work}
	\subsection*{Social influence and collective memory}
	Whether a piece of art or a song will become a success depends not only on the intrinsic quality of the art piece, but also depends on social factors such as social influence \cite{socialanalytis, algorithmburghardt, socialcosimato, socialoremrod, experimentsalganik} and collective memory \cite{memorycandia, lorenzattention}. 
	Popularity rankings are known to affect how people perceive product quality \cite{popularitybiasabdollahpouri}, and so, success in music charts is likely to be self-reinforcing to some degree. 
	In charts, social influence between rankers has also been shown to increase inequality and instability in the rankings \cite{algorithmburghardt, experimentsalganik} increasing the gap between popular and unpopular products~\cite{concentrationcrain}. 
	This effect is similar to the ``winner-takes-all'' effect that has been described in the economic literature in recent years \cite{winnerrobertg}.  
	In the special case of music markets, the winner-takes-all effect predicts that some superstars can capture a significant portion of the market with just a few songs \cite{tailsbrynjolfsson}. 
	A competing theory of the music market is the ``long-tail'' theory which suggests that an increasing number of audiences are turning to less popular artists, reducing the concentration around a small group of prestigious performers \cite{superstarsordanini}.  
	Social influence among rankers has also been shown to increase the unpredictability of markets \cite{experimentsalganik}. 
	In the presence of social influence among rankers, research has found that the best products are unlikely to perform poorly, the worst products are unlikely to perform best, but all outcomes apart from these extremes are possible \cite{tailsbrynjolfsson}. 
	
	In the past decade, social media have been important in shaping trends in music \cite{socialmediasanjeev}. 
	Social networks such as Twitter are used by users to share and spread information, and these social media are also used as marketing tools for products \cite{twitterjansen}. 
	Cosmiato \textit{et al.} \cite{socialcosimato} created a highly accurate model for predicting album positions in Billboard 200 music charts based on the traffic and sentiment from social media posts concerning the albums before and after their release (the model achieved 97\% accuracy in predicting the approximate chart positions of albums).  
	Park\textit{ et al.} found a correlation between users' social media activity and the diversity of their music consumption. 
	The study also established a link between users' music preferences and their socioeconomic status \cite{musicaldiversitypark}. 
	For example, individuals who followed high-profile news media were much more likely to have a higher level of diversity in their music consumption than those who did not. 
	Moreover, users with diverse musical tastes were less likely to follow music-related accounts on Twitter.
	
	Another important factor of product performance is collective memory \cite{memorycandia, memorygarcia}. 
	Researchers have shown that collective memory is constituted by an initial phase of great attention, followed by a gradual decrease in attention (forgetting) \cite{memorycandia, memorygarcia}. 
	Others have shown that collective attention spans are becoming shorter, presumably due to the increasing amount of information available and its quick spreading \cite{lorenzattention}. 
	Digital media have also increased our consumption, leading to faster exhaustion of interest in a given topic \cite{lorenzattention}.
	
	\subsection*{Success in artistic careers}
	Predicting hot streaks and high impact work in the artistic industries is a very active field of research \cite{onehitberg, janosovsuccess, streaksliu}. 
	One important insight from this line of work is that most career success is indistinguishable from random processes \cite{scienceclauset, williamssuccess}. 
	However, Janosov \textit{et al.} showed that there are careers more and less prone to luck \cite{janosovsuccess}. 
	They have indicated that the success of electronic music is greatly influenced by uncertainty, while classical music is more resilient to this factor. 
	Fraiberger \textit{et al.}~\cite{reputationfraiberger} showed that the success of artists is correlated with access to prestigious institutions. 
	An arrow of causality is not established in this work, though, so it remains unclear whether prestigious institutions are good at spotting talent, success is caused by endorsement from prestigious institutions, or neither of these~\cite{reputationfraiberger}. 
	Williams \textit{et al.} \cite{williamssuccess} investigated the productivity of actors and showed that while they could not predict their efficiency (proportion of active years to the career length), other patterns in acting careers could be spotted. First, the most famous actors are more likely to attract new jobs, expressing the “rich-get-richer” phenomenon~\cite{barabasi1999emergence}.
	Second, they also showed that artists’ careers exhibit bursts of activity and inactivity (hot and cold streaks). 
	
	Many researchers have attempted to create models predicting whether a product will be successful. 
	In the music industry, some methods use sonic features of songs to quantify song quality and use this to predict success \cite{onehitberg, bradlowbayesian, successinteriano, successyucesoy}. 
	The results suggest that performers’ portfolios and the features of products play a significant role in shaping creative careers. 
	Bradlow \textit{et al.} took a different direction and tried to create a statistical model for the Billboard Hot 100 ranking. 
	They augmented the chart data with information on the number of previous Billboard Hot 100 songs and whether the song appeared on a movie soundtrack. 
	Attempting to predict sequences of song positions on the chart, their model managed to capture some of the summary statistics of these sequences. 
	The main outcome, however, was demonstrating that it is difficult to estimate the exact chart position-sequences of songs, obtaining only 10\% accuracy in their predictions~\cite{bradlowbayesian}.
	
	\subsection*{Dynamics of rankings}
	Rankings are a way of presenting hierarchical relationships among a collection of elements. 
	They are a popular approach to describing prestige and popularity in arenas such as sports, academia, film, and music \cite{rankinghazelkorn}. 
	Scientists have been studying the statistical properties of rankings for many years \cite{zipfpowerlaw}. 
	One of the essential pieces of research on rankings in recent years was conducted by Iñiguez \textit{et al.} \cite{dynamicsofranking} who investigated the dynamics of charts for diverse systems ranging from university rankings to usage of words in English, and GitHub repositories. 
	These authors provided evidence for universal temporal patterns of rank dynamics and their observations led them to divide charts into two types: open rankings and closed rankings. 
	Open rankings only have stable top positions (meaning that the songs ranked at the top changed their positions rarely), whereas closed ones also have stable bottom positions.
	Moreover, they discovered two types of movements that shape the charts \-- abrupt movements and smooth/diffusive movements. 
	According to this characterization, items make small position changes around their initial rank in charts with smooth/diffusive movements. 
	Iñiguez \textit{et al.}'s mathematical model reproduces many essential features of real-world rankings.
	
	Another important research on rankings has been conducted by Blumm \textit{et al.} \cite{blummrankings}. They have characterized 3 phases in ranking stability: unstable, score-stable, and rank-stable. They also showed that a noise-driven phase transition induces differences in ranking stability.
	
	Whereas Iñiguez \textit{et al.} focused on the dynamics of entire rankings, largely ignoring the intrinsic quality of individual ranked items, Pósfai \textit{et al.} \cite{talenposfai} focused on the correlation between talent and position in society. 
	They showed that while more talent does not necessarily imply a better rank, a sufficient talent difference is needed to overtake someone in a hierarchy. 
	In the paper, Pósfai \textit{et al.} used an artificial Bonabeau hierarchy model. 
	This kind of model determines the rank of an element by its ability to defeat others in pairwise competitions. 
	Using the model, Pósfai \textit{et al.} found that the only way for elements with smaller talent to overtake those with far greater talent is for the better element to be removed (due to ranking rules). 
	However, for smaller differences, talent plays a more substantial role. 
	These rules apply to many systems in the world, such as rankings of scientists, best-seller lists, or sports rankings \cite{talenposfai}.
	
	Analytis \textit{et al.} \cite{analytisrankings} concentrated on revealing the connection between rankings and social influence. They showed evidence that the rank-based systems drive the “rich-get-richer” phenomenon. They developed a framework for studying discrete-time Markov processes where the distribution of future increments depends exclusively on the relative ranking of components, following a “rich-get-richer” assumption.
	
	\section*{Materials and methods}
	\subsection*{Data collection and preprocessing}
	In this paper, we analyze the historical data on the Billboard Hot 100 music chart. 
	We acquired the full chart data through the chart's website, which contains information on the top 100 most popular songs each week, starting August 4\textsuperscript{th} 1958. 
	To acquire archival charts, a date in the format \texttt{yyyy-mm-dd} can be adjusted in a url-address\footnote{The url that can be adjusted to show historical Billboard Hot 100 charts is \texttt{https://www.billboard.com/charts/hot-100/yyyy-mm-dd/}}. 
	The date indicates the first day of the week of the desired ranking. 
	We chose to collect data from Billboard Hot 100, as it is one of the most influential charts in the world, and a chart that allows users to browse through historical data. 
	The website also contains much information that informs us about chart dynamics: Current position, last week’s position, and total weeks on the chart. 
	The data were saved to a CSV file in a comma-separated format.
	The first row contains the column names. 
	The final dataset consists of $333\,887$ rows and ten columns as described in \autoref{tab:dataset}.
	\begin{table}[]
		\footnotesize
		\centering
		\begin{tabular}{|c|c|}
			\hline
			\textbf{Row name}& \textbf{Row description}  \\
			\hline
			\texttt{first\_day\_of\_the\_week} & The first day of the week’s ranking.\\
			\texttt{artist} & Name(s) of the artist(s) performing the song.\\
			\texttt{song\_name} & The name of the song.\\
			\texttt{position} & Position of the song in the current week.\\
			\texttt{last\_week\_position} & Last week’s position of the song (``-'' if the song just entered the chart)\\
			\texttt{peak\_position} & The highest position reached by the song until the current week.\\
			\texttt{weeks\_on\_chart} & The total number of weeks a song has spent on the chart until the current week.\\
			\hline
		\end{tabular}
		\caption{Description of our data set. For every chart position and every week the Billboard Hot 100 was published, our data contain the shown information.  }
		\label{tab:dataset}
	\end{table}
	
	Some entries in the ``artist'' column are collaborations/features between multiple artists. 
	Without loss of generality, these rows are regarded as distinct artists unless specified otherwise.
	
	\section*{Results}
	In the following, we present our findings in four sections. 
	First, we quantify changes in the lifetime of songs on the Billboard Hot 100 chart since its inception on August 4, 1958. 
	Then, we analyze position sequences of individual songs on the Billboard Hot 100 and argue that these ``song trajectories'' can be used to divide songs into distinct song archetypes. 
	In the third section, we analyze the frequency of individual position changes over time. 
	Finally, we focus on the chart positions achieved by different artists over time and compare the performance of ``hitmakers'' to the performance of other artists.
	
	\subsection*{Changes in songs' lifetimes on the Billboard Hot 100 chart}
\begin{figure}
		\centering
		\includegraphics[width=0.75\linewidth]{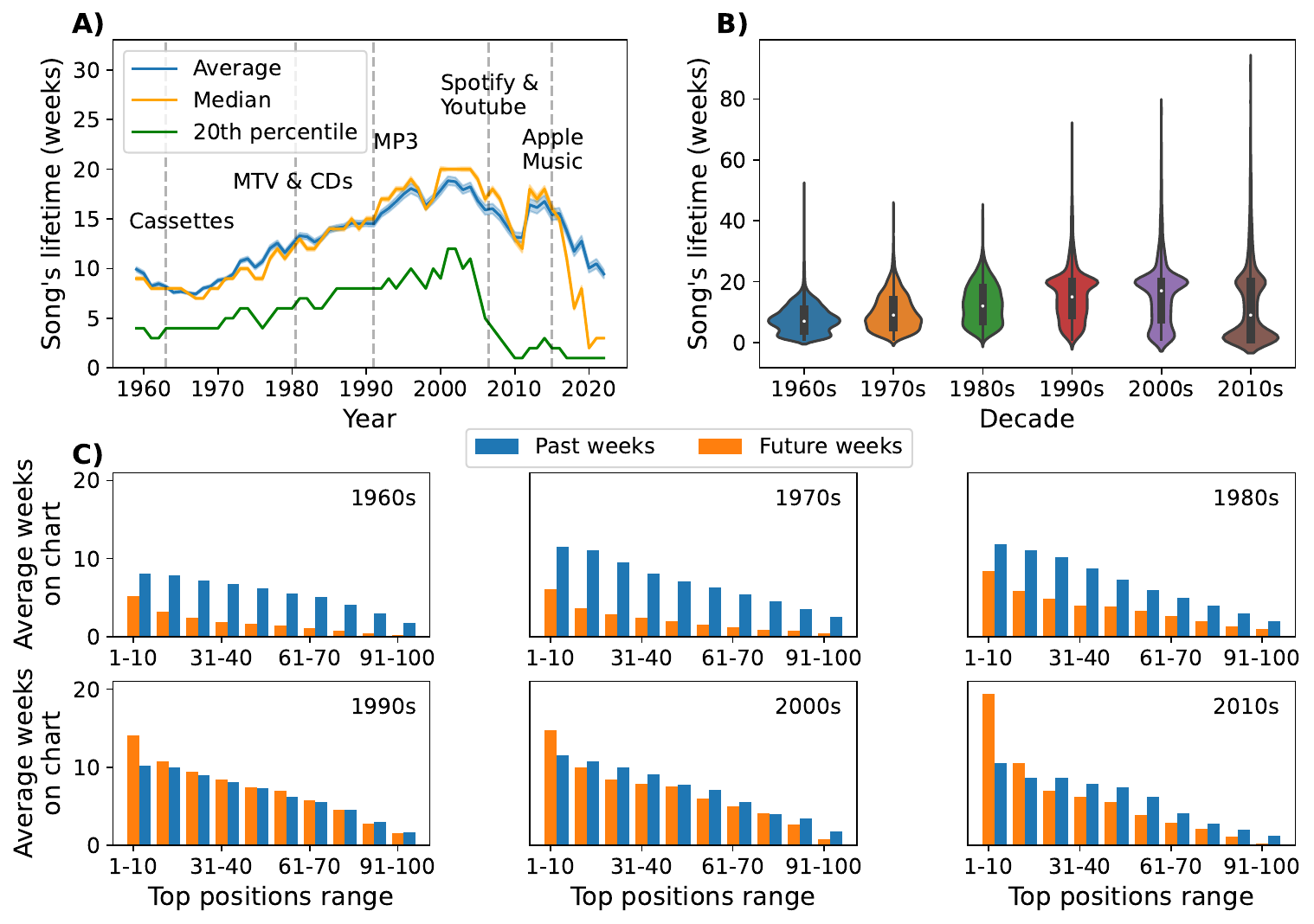}
		\caption{\textbf{Summary of changes in songs' lifetimes.} The figure shows statistics on songs' lifetimes and their changes over time. \textbf{A} The average (blue), median (orange), and the curve indicating the 20th percentile (green) of the lifespans of all songs over time. The shaded areas represent the standard errors of the means. The dashed grey lines indicate some of music history’s most important technology launches. All three plotted song summary statistics have been going up until the 2000s and then started to drop. The median and the 20th percentile have fallen below 5, achieving the lowest values in the chart’s history. The drop in the last 2 years can be partially affected by the songs that have just entered the chart, but their lifetime is yet to be established (they last longer than the end of our data frame). \textbf{B} presents the lifetime distributions as violin plots for different decades. The white dot represents the median, and the dark rectangle shows the interquartile range. The colored areas represent the densities of the distributions. As time progresses, the distributions become thinner and get longer tails. One exception is the formation of clear bumps around week 1 and week 20, which indicate songs that left the chart immediately after entering it, and ``recurrent songs``, which have spent 20 weeks on the Hot 100 and fallen below position number 50 leading to the song's removal from the chart~\cite{chartlegend}. The long tails of the distributions capture the extreme values of the data. In the last decades, the longest-living songs last longer than in previous decades (in the 2010s, reaching four times longer lifetimes than the upper bound of the interquartile range). \textbf{C} The bins represent the average weeks on the chart before the peak (blue) and after the peak (orange). In the first three decades, the blue bars dominated the orange bars, showing that songs quickly disappeared after reaching the peak. Since the 1990s, the bars have been of nearly equal height, except for the top 10, where the average for future weeks is twice higher than for the past weeks. This shows that currently, the best-performing songs linger on for much longer.}
		\label{fig:fig1}
	\end{figure}

	One indicator of a song's success is its lifetime on the charts. \autoref{fig:fig1}A presents change in average, median, and 20th percentile of songs' maximum week on the chart over time.
	Since around 1967, there has been a nearly linear increase in the lifetime of songs in all three of these summary statistics. 
	The last three decades show a decrease in the values, especially drastic for the 20th percentile.
	
	The drop in lifetimes has been the dominant trend since the mid-2000s when song lifetimes were 20 weeks, twice the average lifetime in the first week of 1959. 
	Since this peak in song lifetime, the average song lifetime has returned to the value it had at the chart's inception. 
	Similar trends can be spotted in the median. 
	Until 2015, the median lifetime of songs was almost indistinguishable to the average lifetime of songs. However, starting around 2015, the median lifetime has fallen significantly below the average lifetime, indicating a more left-skewed distribution of song lifetimes. 
	Nowadays, around half of the songs on the chart have a very short lifespan, lasting only a few weeks on the chart.
	These songs have difficulty maintaining their places in the charts, whereas other songs remain on the charts for months and years.
	
	\autoref{fig:fig1}B shows violin plots of lifetime distribution that support the observations so far. 
	The song lifetime distributions in the first three decades are quite narrow, with only a few songs achieving more than 20 weeks on the Billboard Hot 100 chart in these decades. 
	However, the distributions get longer right-tails in the following decades. 
	In the 2010s, the song lifetime distribution has peaks around 5 and 20 weeks, and the longest lifetimes are over twice as high as for the first decades. 
	The 5-week peak of the distribution indicates that many songs disappear from the charts quickly, whereas the 20-week peak indicates that many other songs last for longer than what was the norm in earlier decades (the 20-week peak itself is due to a chart rule that we describe in more detail in the caption of Figure~\ref{fig:fig1}). 
	The median, represented by a white dot, shows the same trend as in the previous plot (\autoref{fig:fig1}A): 
	It has been increasing till the 2000s and dropped in the 2010s. 
	On the other hand, the interquartile range (black rectangle) has been getting wider; this can be understood as an increase in the variability of the data.
	
	At some time during its time on the chart, a song reaches its peak position. 
	Before the song reaches its peak position for the first time, it can be described as being in an ascending phase of its chart lifetime. 
	The ascending phase is followed by a descending phase. 
	\autoref{fig:fig1}C gives us a better understanding of how song lifetimes have changed on the Billboard 
	Hot 100 over the decades, we plotted the average number of weeks songs spent on the chart before and after reaching their top position. 
	The figure shows that the average future weeks on the chart have decreased over the last three decades for all positions except those that reached the top 10. 
	Conversely, songs that reached the top 10 have experienced an increase in average lifetime in each of the decades following the chart's creation, now reaching average lifetimes almost four times longer than in the 1960s. 
	
	In the top 3 panels of \autoref{fig:fig1}C, blue bars dominate orange bars. 
	This means that, during the first three decades, songs took longer to secure their peak position than they did to disappear from the chart after they reached the peak. 
	The tendency reversed for almost all peak positions in the 1990s. 
	Only the worst-performing songs spent more weeks on the chart before peak position than after in this decade. 
	Nowadays, the best-performing songs spend, on average, twice as long time on the chart after their peak as before it. 
	However, entries at the worst ranks (bottom 20) peak and disappear faster than in the first decades of the Billboard Hot 100 chart. 
	These results illustrate the intensifying inequality between certain groups of songs.
	
	\subsection*{Song trajectories and archetypes}
	So far, our results have indicated that there has been a change in how songs perform on the Billboard Hot 100 over the last 60 years: 
	Songs that do not peak at high positions are more likely to disappear from the chart very quickly, whereas the best ones linger on the chart for much longer. 
	The observation that some songs disappear more quickly whereas others stay for longer presents the question of whether songs can be divided into inherently different classes. 
	To investigate whether such discrete song \textit{archetypes} exist, we used the $k$-means clustering algorithm to group songs based on the following observables of song chart trajectories: maximum weeks on the chart, weeks before the top position is reached for the final time, weeks after the top position is reached for the final time, starting position on the chart, and end position on the chart. 
	We chose these features, as they represent key characteristics of the song trajectories in our first analysis.
	Based on the Silhouette scores obtained when asking the algorithm to find different numbers of clusters, $k$, in the song trajectory data, we found that all songs from the Billboard Hot 100 chart can be clustered into $5$ song archetypes. 
	We show the statistics on their features in Table~\ref{tab:clusterstats}. Based on the 75th percentiles of weeks on the chart, first and last positions, we found distinct characteristics in the clusters:
	\begin{itemize}
		\item In Cluster 0 the 75th percentile of weeks on chart is significantly smaller than for the other groups.
		\item Cluster 2 has a low 75th percentile of end position.
		\item Cluster 3 has the lowest 75th percentile of first position.
		\item Cluster 4 has the highest 75th percentile of weeks on chart.
	\end{itemize}
	Based on the visual and statistical characteristics of the songs in the five clusters,  we give the song archetypes names as shown in Table~\ref{tab:clusterstats}. 
	
	\begin{table}[]
		\centering
		\footnotesize
		\begin{tabular}{|c|c|c|c|c|c|}
			\hline
			\textbf{Song archetype }& \textbf{Cluster}&\textbf{\begin{tabular}[c]{@{}c@{}}Weeks on \\the chart\end{tabular}} & \textbf{\begin{tabular}[c]{@{}c@{}}First\\position\end{tabular}} & \textbf{\begin{tabular}[c]{@{}c@{}} End\\position\end{tabular}} & 
			\textbf{\begin{tabular}[c]{@{}c@{}}Qualitative trajectory\\characteristics\end{tabular}}
			\\ \hline
			\begin{tabular}[c]{@{}c@{}}Brief\\ songs\end{tabular}  &0      &  9 & 96  & 97 & \begin{tabular}[c]{@{}c@{}}Last only a few weeks\\ on the chart\end{tabular}\\ \hline
			\begin{tabular}[c]{@{}c@{}}Climbing\\ songs\end{tabular}  &1  &  20 & 91 & 98 & \begin{tabular}[c]{@{}c@{}}Start at low chart positions, \\ climb up to their top positions, \\and then slowly decay.\end{tabular}\\ \hline
			\begin{tabular}[c]{@{}c@{}}High-end\\ songs\end{tabular} &2 &  18 & 93 & 56 & \begin{tabular}[c]{@{}c@{}}Leave the charts quickly  \\ after attaining high top \\positions.\end{tabular} \\ \hline
			\begin{tabular}[c]{@{}c@{}}High-start\\ songs\end{tabular} &3 & 20& 43& 95 &\begin{tabular}[c]{@{}c@{}}Enter the charts at\\ high start positions.\end{tabular}\\ \hline
			\begin{tabular}[c]{@{}c@{}}Long-lasting\\ songs\end{tabular} &4& 40& 90& 49&\begin{tabular}[c]{@{}c@{}}Stay on the chart\\ for a long time.\end{tabular}\\ \hline
		\end{tabular}
		\caption{Summary of clusters of song trajectories found by applying the $k$-means clustering algorithm to summary statistics of all song trajectories in the history of Billboard Hot 100. We find $5$ clusters of song trajectories and list the 75th percentiles of $3$ features used in identifying clusters: Number of weeks on the chart, first position on the chart, and end position on the chart. Based on these features, we give the clusters an ``Archetype Name''. ``Long-lasting songs'' hold the record for the highest number of weeks on the chart, whereas cluster 2 has high end position. Cluster 0 has the smallest weeks on the chart, cluster 3 has the lowest first position. These values give a better overview of the possible archetypes.}
		\label{tab:clusterstats}
	\end{table}
	
	\begin{figure}
		\centering
		\includegraphics[width=0.75\linewidth]{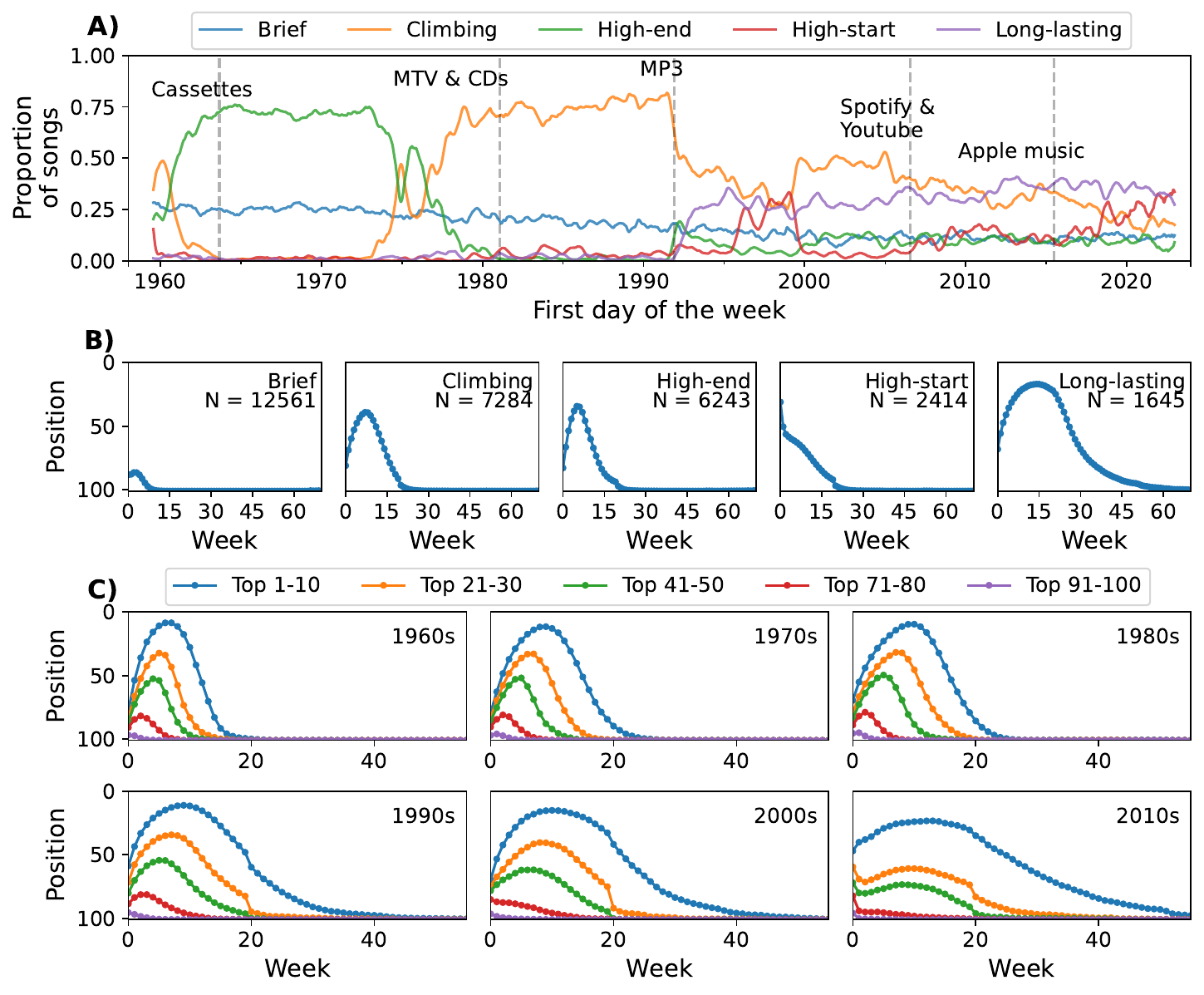}
		\caption{\textbf{Trajectory and archetype analysis.} Key aspects of songs' trajectories and their evolution over time. \textbf{A} The proportion of songs from each proposed archetype over the years. The results were normalized with a half-year rolling window. The groups are brief songs (blue), climbing songs (orange), high-end songs (green), high-start songs (red), and long-lasting (purple). The dashed grey lines indicate the timing of some of the most important technological innovations in music listening history. First, high-end songs dominated until the mid-1970s when the climbing flow took over. Since the 1990s, the climbing songs have become less dominant, now competing with long-lasting, high-start, and the revitalized high-end songs. Over the course of the chart’s history, the prevalence of short songs has steadily declined from 30\% to approximately 10\%. \textbf{B} Normalized trajectories and the number of songs in each cluster. The trajectories were calculated by averaging the positions of all songs in each week following their first appearance on the chart. The plots summarize the visual aspects of detected archetypes. One thing that stands out is that the clusters are not evenly sized. The brief songs group is almost two times bigger than the second-largest, climbing flow. \textbf{C} Normalized trajectories of songs grouped by top position of the songs. Grouped together are songs whose top position were in top 10 (blue), positions 21-30 (orange), positions 41-50 (green), positions 71-80 (red), and positions 91-100 (purple). Only a few position ranges were chosen for better readability of the figure. The missing intervals lie between the visible ones (e.g., the 11-20 curve lies between the top 1-10 and the top 21-30). The songs that disappear from the charts have the rest of their trajectories filled with 101. As time passes, the average curves widen and start higher at week zero. There is also a visible drop in the 20 weeks, explained in the Billboard Hot 100 as the “recurrent” status \cite{chartlegend}}
		\label{fig:fig2}
	\end{figure}
	
	In \autoref{fig:fig2}A, we investigate the prevalence of song archetypes throughout the history of the Billboard Hot 100 chart. 
	We find clear evidence of changes in song trajectories in the past 60 years. 
	In the first two decades, almost 80\% of the songs are of the ``high-end'' archetype. 
	As we showed in the previous section, these songs disappear quickly from the charts after reaching their peak positions. 
	The second-largest group in the first decades is ``brief songs''. 
	In the middle of the 1970s, the ``high-end'' trajectory archetype became less prevalent on the chart, and ``climbing songs'' became the new dominating archetype. 
	In the 1990s, the proportion of climbing songs decreased, and a few new groups rose: the ``long-lasting'' and ``high-start'' archetypes. Moreover, the high-end pieces started to reappear after a break around the 1980s-1990s. 
	The brief songs have gradually decreased since the 1980s but still account for around 10\% of all songs today. 
	The long-lasting and high-start songs were the most influential in the most recent years. 
	Combined, they account for approximately 60\% of all songs. 
	
	\autoref{fig:fig2}B illustrates the mean position of a song during each week following its appearance on the chart conditional on the song's archetype. 
	The long-lasting songs stay on the chart for longer, and the high-start songs start more elevated than others. 
	High-end songs disappear quickly following peak positions compared to the slower disappearance of climbing songs. 
	Finally, the mean trajectory of brief songs has the lowest height and width. 
	$N$ indicates the number of songs in each archetype cluster. 
	The largest group is “Brief songs,” with over 12000 songs, and the smallest is “Long-lasting,” with only 1645. 
	(We note that $N$, the number of songs in each cluster, is different from the proportion of songs on the chart which we plot in \autoref{fig:fig2}A; though few songs are long-lasting (low $N$), a large percentage of songs on the chart can be of this archetype every week because each songs stays on the chart for a long tme.) 
	\autoref{fig:fig2}B shows that the clusters are distinguishable and sensible propositions for song archetypes.
	
	\autoref{fig:fig2}C, shows average trajectories conditional on song peak position for each decade following the inception of Billboard Hot 100. 
	The figure panels show that the averaged song trajectories have flattened out over time, again demonstrating that some songs now stay on the chart for longer. 
	The flattening happens very quickly for the worst positions. 
	However, it can be seen that all the averaged curves in the 2010s start higher than in the previous decades indicating that songs begin their chart trajectories at better positions. 
	That the peak of averaged trajectories in more recent decades is lower than the peak positions of the songs that contribute to the averaged trajectories indicates that the songs reach their peak position at more different times than used to be the case~\cite{juul2021fixed}.
	
	\subsection*{Changes in jumps}
	We now turn our attention from trajectories to the weekly chart position jumps taken by songs.
	\begin{figure}
		\centering
		\includegraphics[width=0.75\linewidth]{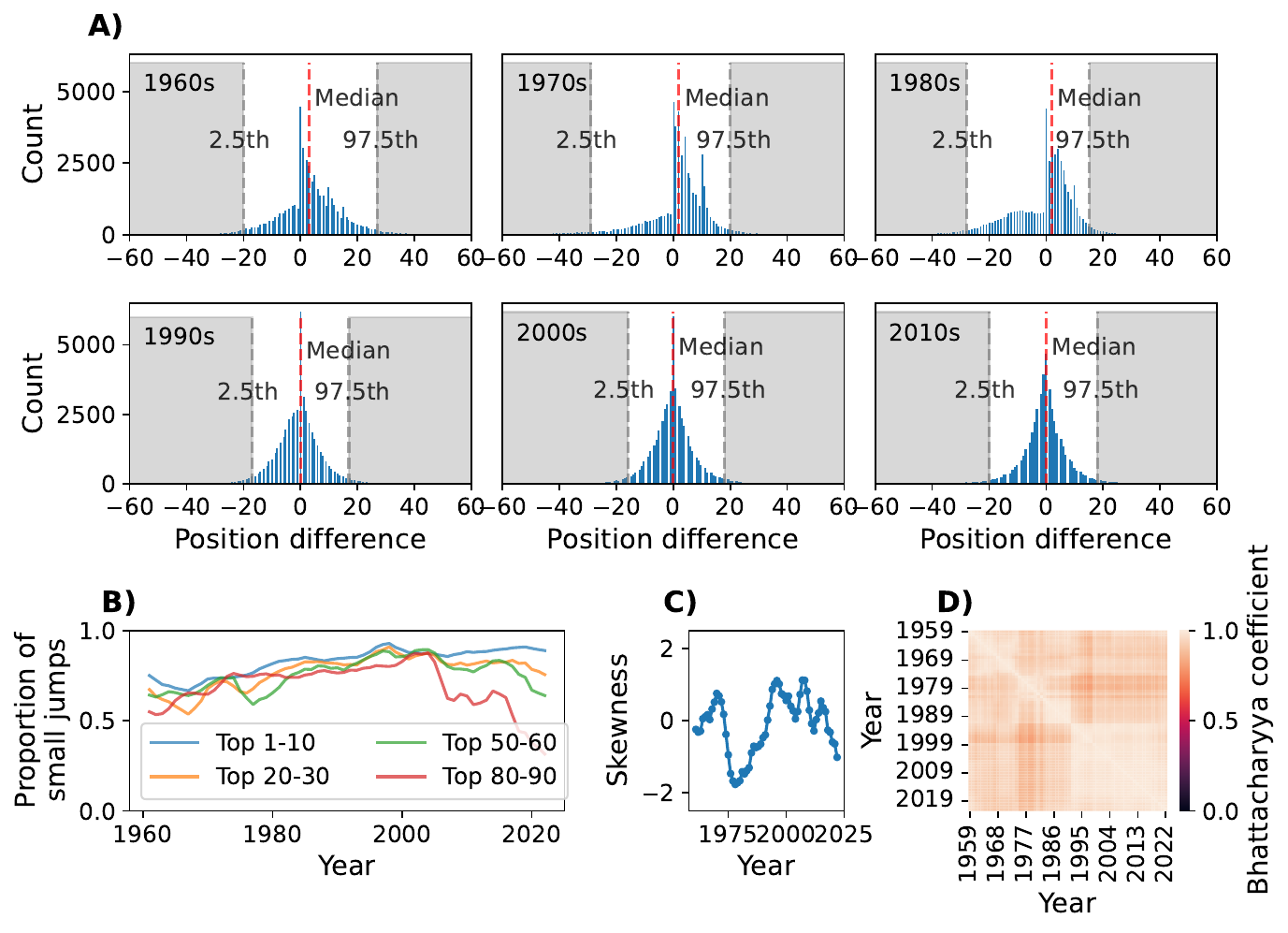}
		\caption{\textbf{Weekly rank changes over time.} \textbf{A} Histograms of weekly position changes over decades. The red dashed lines represent the median of the samples. The grey dashed lines represent the samples’ 2.5th and 97.5th percentiles, respectively. The shaded grey areas show the elements smaller than 2.5\% of the data and greater than 97.5\% of the data. During the first three decades, the median skewed towards the right, but it ultimately shifted towards zero in later years. Moreover, the distributions got more symmetrical with time, whereas before they were concentrated on positive position differences. \textbf{B} Proportion of small jumps for songs lying in 1-10 (blue), 20-30 (orange), 50-60 (green), and 80-90 (red) songs. The values have been smoothed with a moving average of 3 years. We show these 4 groups of songs for ease of readability and to align with the previous analysis of the best-ranking positions. The overall proportion of small jumps started to drop around the 2000s, but for the top 10, it stabilized at around 80\%. For the lowest positions, the proportion drastically dropped to around 30\%. \textbf{C} Skewness of yearly position change distributions. The plots have been smoothed with a rolling average of 3 years. The overall skewness has been changing a lot throughout the years. There are fewer fluctuations in the center values. In the first half, the majority of them register values below 0, while in the second half, they record values close to 0. \textbf{D} Heatmap of similarities between yearly position change distributions. The similarities were calculated as Bhattacharyya coefficients between distributions from different years. The colors indicate the similarity between distributions from one year (x-axis) and another year (y-axis). The lightest colors exhibit great similarity, while the darkest do the contrary. The biggest distribution overlap can be spotted between the years in the last three decades (bottom right corner), while the largest disparities are found between the 70s and the period after the 80s, and the 90s and the years before them.
		}
		\label{fig:fig3}
	\end{figure}
	\autoref{fig:fig3}A shows that the frequency of weekly position changes over decades. 
	The median shifted to the right for the first three decades, stabilizing at 0 since the 1990s. 
	Another difference is the distribution shape: For later decades, the distributions are increasingly symmetric. 
	This may imply that nowadays, songs move more smoothly on the charts, gradually increasing and decreasing their positions. 
	To test this theory, we define a small jump as an absolute weekly position change smaller than 10 positions. 
	Otherwise, a rank difference is considered a long jump. 
	We show the proportions of these measures for all songs and different top position ranges in \autoref{fig:fig3}B. 
	The figure shows that the prevalence of small jumps has, on average, increased over time but started dropping around the 2000s for songs outside of top 10 positions. 
	For the worst-performing songs, the proportion has fallen to roughly 30\%, compared to 80\% for songs that peaked at the top 10.
	
	In \autoref{fig:fig3}C we quantify the skewness of position change distributions for different decades with the Fisher-Pearson coefficient. 
	We show the plots in \autoref{fig:fig3}C. 
	There is a significant trough in the skewness in the 1970s-1990s period. 
	Since then, the overall skewness has been fluctuating around zero before dropping again in recent years. 
	Distributions with negative skewness have a heavier right tail, with more values concentrated on the right side. 
	This indicates that most songs maintain a stable climbing-up flow, with a few experiencing drastic jumps and drops.
	
	In \autoref{fig:fig3}D we show the similarities between position change distributions from different years. The heatmap reveals a few patterns - first, the diagonal forms the brightest line, indicating nearly identical distributions between consecutive years. This shows that the change in the distributions over the years has been smooth. The last 3 decades form a light square in the bottom right corner. This shows that there hasn't been much change in the distributions in the last 30 years. Lastly, there are 2 rectangles of slightly darker color (lower coefficient values). While the coefficient is still relatively high (around 0.8), we can see disparities between the 70s and the period after the 80s, and the 90s and the years before them. Overall, the heatmap resembles the results from the \autoref{fig:fig3}A. 
	
	\subsection*{Hitmakers and artist performance}
\begin{figure}
		\centering
		\includegraphics[width=0.75\linewidth]{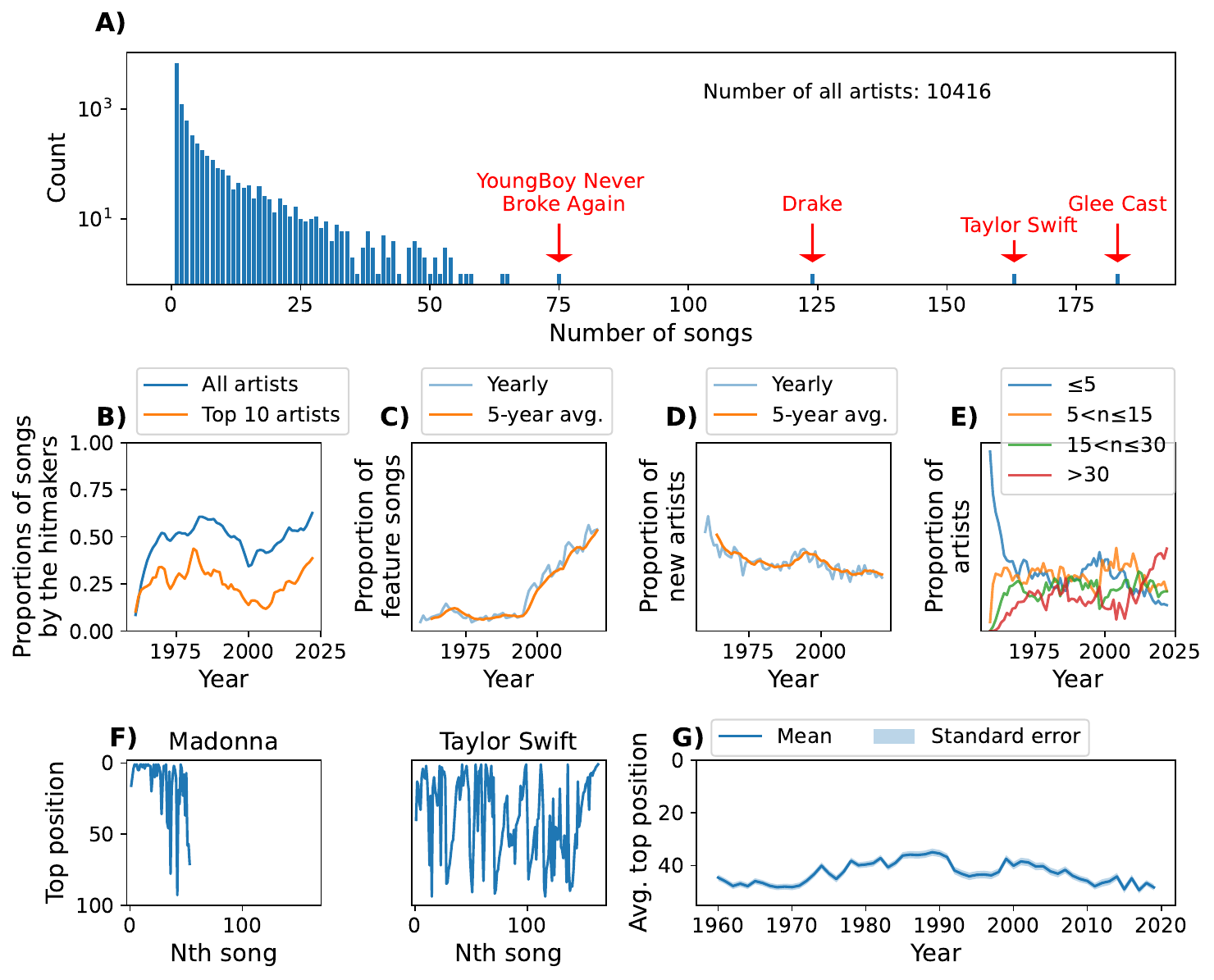}
		\caption{\textbf{Artists on the Billboard Hot 100 chart.} The figures show various statistics on artists on the chart. \textbf{A} Distribution of the number of songs on the chart per artist. The red arrows point to the outliers with very high song count. The counts have been presented on a logarithmic vertical scale, as the vast majority of artists that get a songs on the Billboard Hot 100 chart achieve this feat for only 1 or 2 songs (80\% of all artists that get on the chart). The top 3 artists (Glee Cast, Taylor Swift, and Drake) are far from the rest of the distribution. \textbf{B} Proportion of hitmakers on the charts relative to all artists (blue) and those who achieved the top 10 (orange) over the years. The figure was smoothed with a 3-year moving average. In the 2000s, only 10\% of the artists in the top 10 and 40\% overall had more than ten songs on the charts. Today, hitmakers account for 60\% of all artists in the charts and over 40\% of the top 10. \textbf{C} Proportion of features/collaborations over time. The blue curve presents the actual values, and the orange is smoothed with a 5-year window. The ratio was more or less steady till the 1990s, fluctuating around 10\%. However, since the late 1990s, it has rapidly increased, achieving around 50\% in 2022. \textbf{D} Proportion of new artists (artists appearing for the first time on the weekly chart) over the years. The proportion of new artists without features is decreasing with time. Nowadays it is around 30\%. \textbf{E} Proportion of individual artists with different numbers of songs. The different ranges of the number of songs are:  $\le 5$ (blue), $5<x\le 15$ (orange), $15<x\le30$ (green), and more than $30$ (red). Plots have been smoothed with a 5-year moving average. The ratio of the smallest number of songs (blue) is naturally very high at the beginning of the chart’s history. It has dropped to around 20\% in the last years. The medium values (orange and green) have been fluctuating between 20\% and 40\%. The red curve, which represents artists with over 30 songs on the Billboard Hot 100, has visibly increased in the last two decades. \textbf{F} Top positions of songs by Madonna and Taylor Swift. The x-axis represents the nth song of the artists, and the y-axis its top position. Many of Madonna's songs reached the top 1 and they rarely achieved worse positions than 50. For Taylor Swift there are more frequent fluctuations in their top positions - some of her songs did not even reach top 80. \textbf{G} Average top position of hitmakers' songs over the years. Their songs performed, on average, slightly better in the 1980s-2000s period, compared to nowadays. In recent years the mean position of hitmakers' songs has dropped to around 50.}
		\label{fig:fig4}
	\end{figure}

	So far we have examined how song trajectories have changed over time, and how trajectories differ for songs reaching high and low peak chart positions. 
	Now we turn our focus to artists and their chart performance across the decades. 
	\autoref{fig:fig4}A shows the distribution of the number of songs artists have placed on the Billboard Hot 100 throughout the chart's history (note the exponential vertical axis).
	\autoref{fig:fig4} shows that the vast majority of performers have no more than a few songs on the prestigious chart throughout their whole career. 
	Only 5\% of artists achieve to get more than 10 songs on the chart. 
	With that in mind, we classify an artist as a hitmaker if they have had over 10 songs on the chart in the past (our results are robust when changing this threshold, e.g. to 5 and 20 songs).
	
	\autoref{fig:fig4}B shows that the proportion of songs on the Billboard Hot 100 created by hitmakers has changed nonlinearly over the years. 
	The proportion of chart songs made by hitmakers first increased until the 1980s, then decreased for a few years, before proceeding to rise again in the 2000s. 
	In the 2000s, 10\% of the artists in the top 10 and 40\% overall had had more than ten songs on the chart. 
	Today, hitmakers account for an all-time high of 60\% all songs in the charts and almost 40\% of the songs in top 10.
	
	Another interesting change is the tendency for artists to collaborate on hit songs. 
	\autoref{fig:fig4}C shows the proportion of collaborative songs over the years. 
	The proportion of features and collaborations was more or less steady till the 1990s, fluctuating around 10\% before it increased rapidly. In 2022, around 50\% of chart songs are features and collaborations. 
	Focusing on artists that appear on the chart for the first time, \autoref{fig:fig4}D shows that there has been a gradual decrease in the proportion of new artists on the chart over time, except for a small spike around the 1990s. 
	The proportion of new artists on the chart is now approximately $30\%$, down from 60\% in the 1960s.
	
	\autoref{fig:fig4}E provides a more detailed picture of how many hits artists on the chart have had in the past. 
	The number of artists with more than 30 past hit songs has increased visibly since the 2000s, peaking at around 40\% in 2023. 
	If artists stick around for longer, it is to be expected charts should be occupied by artists with an increasing number of past hits. 
	However, the growth of the prevalence of multi-hit artists on the charts is not monotonous over the decades, and especially the prevalence of artists with more than 30 hits has been accelerating visibly since the 2000s.
	
	Finally, we focus on how the performance of hitmakers has changed over the decades. 
	In \autoref{fig:fig4}F we show careers of two famous pop artists: one from the late 1980s \-- Madonna \-- and one from the 2010s \-- Taylor Swift. 
	While Madonna had a very good start, with most of her first songs getting to top 20, her songs' peak positions started to fluctuate heavily in later years. 
	For Taylor Swift, peak position has fluctuated since she started her career, with some of her songs reaching number one, but other struggling to achieve the top 90. 
	These two examples could suggest that hitmaker songs have become less certain to reach top chart position in recent decades. 
	In \autoref{fig:fig4}G, we test this hypothesis by computing the average top position of hitmaker songs throughout the years. 
	Although the variations are not dramatic, \autoref{fig:fig4}G shows that there were times in music history, when the best-performing artists achieved higher positions on average. 
	Today, the hitmaker peak position average is at an all-time low and seems to be decreasing, indicating that new hitmaker singles do not necessarily attain critical acclaim on the chart. 
	This effect is partly caused by the fact that a lot of their non-single songs end up in the rankings. With limited airplay slots, non-singles getting less exposure (and may be songs that are deemed inherently less likely to please a large audience), and so, the non-singles are more destined to fail. Evidence of this effect can be seen for example in the songs' listener counts. In The Weeknd's “After Hours” album Last.fm statistics\footnote{\href{https://www.last.fm/music/The+Weeknd/After+Hours}{https://www.last.fm/music/The+Weeknd/After+Hours}}, we can see that the four singles “Blinding Lights” (1,653,759 listeners), “Save Your Tears” (1,243,019 listeners), “Heartless” (902,088 listeners) and “After Hours” (811,096 listeners) have a significantly higher number of listeners than the non-single songs, e.g., the lowest performing “Until I Bleed” with only 340,107 listeners. Moreover singles, in general, perform better on the charts - e.g., “Blinding Lights” achieved the top 1 on Billboard Hot 100 and stayed on the chart for about 90 weeks, while “Until I Bleed” peaked at 80th position and lasted for only 1 week\footnote{\href{https://www.billboard.com/artist/the-weeknd/chart-history/hsi/}{https://www.billboard.com/artist/the-weeknd/chart-history/hsi/}}.

	\section*{Discussion}
	
	The Billboard Hot 100 chart is one of the most important music charts in the world. 
	Our analysis of song performance on the chart throughout the chart's history indicate several changes in the dynamics of songs and artists on the chart. 
	Compared to the first decades, song lifetimes are more skewed: some songs stay on the chart for many months, whereas the majority disappear very shortly after they appeared. 
	The average and median song lifetimes have increased since the creation of the chart, and whereas songs used to slowly climb the chart and then disappear quickly after reaching their peak position, many songs now spend more time on the chart after reaching their peak position than they did before reaching the peak. 
	Today's chart has fewer new artists making their debut on the chart, more songs that are collaborations or features, and more spots occupied by established artists that we refer to as hitmakers. 
	All these findings document how a range of changes in the pop music charts,  make it more difficult to make a hit.
	
	The culmination of our findings was the identification of archetypes of trajectories over time. 
	We found five different typical paths of songs on the charts and showed their prevalence over time. 
	The results show that archetypes have been more and less dominant on the chart in different time periods, demonstrating how chart dynamics have changed over time. 
	Until the 1970s, songs would quickly disappear from the charts after reaching their peak. 
	Then followed a period of time that lasted until the 1990s, where songs would climb from the bottom of the ranking to their peak and then diffuse slowly to the bottom positions. 
	The most-recent decades show the emergence of long-lasting songs and songs that start at the top positions of the chart. 
	Our results could be explained by the recent findings on attention acceleration. 
	Due to the increasing speed and spread of information through the Internet, cultural items gain a vast amount of attention quickly in the beginning, but the attention spans last shorter \cite{lorenzattention}.
	
	Some of our results allude to increasing competition in the music industry, and the chart data hint that artists and record companies have developed new strategies to increase performance on the charts. 
	Strategy changes are especially visible in the increasing number of collaborations and features on the chart, which could be a way for creators to mobilize multiple fan bases to boost listener numbers. 
	Another sign that strategies may have changed is the observation that hitmakers now occupy more spots on the chart, artists often having multiple singles on chart top positions at the same time.  
	That the music industry has changed in recent years is mirrored by findings from other studies \cite{hierarchyclauset, prestigethornton}. 
	The impact of online information spread and advertising is increasing significantly. 
	The most prestigious artists are more likely to be endorsed and gain the most out of advertisements \cite{hierarchyclauset, prestigethornton, skewnesshendricks}. 
	In contrast, albums from less famous artists tend to sell below their potential due to a lack of proper promotion. 
	In summary, the industry seems to be more biased toward the superstars \cite{skewnesshendricks}.
	
	Two competing and seemingly contradicting theories concerning the skewness of the music market have emerged in the literature in recent years. These theories can be referred to as the ``winner-takes-all'' theory and the ``long-tail'' theory \cite{superstarsordanini}. 
	The supporters of the winner-takes-all theory say that some superstars can capture a significant portion of the market with just a few songs. 
	On the contrary, the long-tail effect states that more and more audiences switch to less popular artists, decreasing the concentration around the most famous performers. 
	Interestingly, in our work, there is evidence that both effects are present in modern music. 
	Our finding that songs debut at better positions and have longer chart lives may be due to the winner-takes-all superstar phenomenon. 
	Conversely, the “long-tail” effect can be spotted in the increased traffic in the chart’s bottom positions and shorter lifetimes for poor-performing songs. 
	
	Finally, an interesting aspect of the Billboard Hot 100 historical data is the absence of accelerated public attention. 
	In recent years, researchers have documented accelerated public attention across domains: 
	From social media hashtag trends to book sales, movie box-office sales, scientific publications, and more, hits are coming and going at a higher pace~\cite{lorenzattention}. 
	In the Billboard Hot 100 chart, however, we see some hits dominating the chart for much longer periods than used to be the case. Investigating this difference between music and other domains is a promising direction for future research.
	
	\subsection*{Limitations}
	
	The music industry is a complex market - fast-paced, competitive, and highly dependent on consumers. 
	Our analysis of the historical Billboard Hot 100 chart suggests that it may have gotten harder to create hits over the years. 
	While the music charts provide a lot of valuable information on the changes in the music industry and music consumption, there are many different aspects of songs. 
	These include live performance sales, sonic features, social media traffic, etc. 
	This means that chart data tell us about only a part of a song's success. 
	Other relevant data to quantify song success could include sonic features of songs, historical information on listening counts, sales of the singles/albums, social media interactions, and more. 
	
	Moreover, it is a clear limitation that the data only includes the ranking of 100 different songs every week. For an even more comprehensive analysis, key figures for a wider range of tracks should be followed over time.
	
\section*{Availability of data and materials}
The datasets collected and analysed during the current study are available in the following github repository, \texttt{ https://github.com/martalech/pop\_charts\_dynamics/ }. 

\section*{Competing interests}
The authors declare that they have no competing interests.

\section*{Author's contributions}
ML designed the research, acquired the data, performed data analysis and wrote the paper.  SL designed the research, supervised the project and wrote the paper. JLJ designed the research, supervised the project and wrote the paper.

\section*{Acknowledgements}
JLJ thanks Alice Nadeau for inspiring conversations that led to the creation of this research project. JLJ's work presented here is supported by the Carlsberg Foundation, grant CF21-0342.


\end{document}